\documentstyle{article}
\newcommand{\n}{\nonumber}
\newcommand{\bea}{\begin{eqnarray}}
\newcommand{\eea}{\end{eqnarray}}
\newcommand{\be}{\begin{equation}}
\newcommand{\ee}{\end{equation}}
\newcommand{\ph}{\phi}
\newcommand{\fr}{\frac}
\newcommand{\tl}{\tilde}
\newcommand{\sm}{\sum_n}
\newcommand{\ld}{\ldots}
\newcommand{\gm}{\Gamma}
\newcommand{\sq}{\sqrt}
\newcommand{\md}{\mid}
\newcommand{\al}{\alpha}

\title{Coherent States\\ of\\ Non-Linear Algebras:\\ Applications to Quantum Optics.}
\author{V. Sunilkumar, B. A. Bambah\thanks{e-mail:
bindu@uohyd.ernet.in, also from Dept. of Mathematics, Panjab University, Chandigarh.}, R. Jagannathan$^2 $\thanks{e-mail jagan@imsc.ernet.in},
P. K. Panigrahi\thanks{panisp@uohyd.ernet.in} and V.
Srinivasan\thanks{vssp@uohyd.ernet.in}\\
School of Physics,
University of Hyderabad,
Hyderabad, Andhra Pradesh,
500 046, India,\\
$^2$Institute of Mathematical Sciences, Taramani, Chennai, India}
\date{}

\begin{document}
\maketitle

\begin{abstract}
We present a general unified approach for finding the coherent states of 
 polynomially deformed algebras such as the quadratic and Higgs algebras,
which are relevant
for various multiphoton processes in quantum optics.
We give a general procedure to map
these deformed algebras to appropriate Lie algebras.
This is used, for the non-compact cases, to obtain the annihilation operator eigenstates,
 by finding the canonical
conjugates of these operators.
 Generalized coherent states, in the
Perelomov sense, also  follow from this construction.
This allows us to explicitly construct coherent states associated with various quantum optical systems.
\end{abstract}

\large
\section{Introduction}

Till recently, in quantum optics, only linear Lie algebras 
have been used to give mutiphoton coherent (including squeezed) states \cite{kla,perv}.
It is well-known that if we have bilinear Hamiltonians for two mode radiation
fields characterized by operators $a$, $b$, $a^{\dag} $ and $b^{\dag} $,
then the simplest types of coherent states that can be constructed are the product states $|\alpha>|\beta>$ where $|\alpha>$ and $|\beta>$ are the eigenstates
corresponding to $a$ and $b$ respectively.
However, if the system has  an added symmetry or conservation law, then,
 a  set of coherent states  restricted by the extra symmetry can be constructed, by a suitable projection from the ordinary product states.
Examples of such coherent states include the coherent states of a radiation field
with arbitrary polarization such that $a^{\dag} a + b^{\dag} b $ is
conserved. Here, the symmetry algebra is $ SU(2) $, and  the corresponding  states
are the $SU(2)$ coherent states \cite{atcs}. 
In the frequency
conversion of photons of a given frequency $\Omega$ into two photons
of frequencies $\omega _a $ and $\omega _b $, when the two photons are created
or destroyed together such that the operator $Q =a^{\dag} a -b^{\dag} b $
is conserved, the relevant states are 
 the `pair coherent states' or the SU(1,1) `Barut-Girardello (BG)' states.
 \cite{Agarwal,Barut,bhaum}.
 The symmetry algebra in this case is $SU(1,1)$, defined by
$K_{+} = a^{\dag} b^{\dag} ,~~~~K_- = ab ,~~~~ K_0 =
\fr{1}{2} (a^{\dag} a + b^{\dag} b +1)
$
and the Casimir operator $ C=\fr{1}{4} (1 -(a^{\dag} a - b^{\dag} b )^2 ) =
\fr{1}{4} (1- Q^{2} )$.
The coherent states 
are the
simultaneous eigenstates of $ K_- $ and $C$.
Other coherent states  are the SU(1,1)
 Perelomov \cite{perv}
states, 
$
\md \lambda >= e^{(\lambda K_+ - \lambda ^* K_- )} \md 0,0 >
$
,which are the `Caves-Schumaker' states that represent two-mode squeezing \cite{Caves}.
A third set of eigenstates
$
A\md \psi > = (u K_0 + \nu K_- + c K_+ ) \md \psi >= \tilde{\lambda }\md \psi >
$
called algebraic coherent states have also been constructed by Agarwal\cite{satyaprakash}, Shanta et.al \cite{cs}, Trifonov \cite{trif} and
others \cite{manko,brif1}.
The BG states can be produced by a dissipative process governed by
Hamiltonians of the form $ H = H_0 + \omega _1 ab +
\omega_{1}^{*}a^{\dag} b^{\dag} =H_0 + H_{int} $, as the steady states of the
master equation
$\fr{d\rho}{dt}= [\rho, H_{int}]$. 
The $SU(1,1)$ Perelomov states represent the time evolution of the
same Hamiltonian.
Thus, the corresponding Lie algebraic structure has proved instrumental
in studying the quantum optical properties of two mode radiation fields.

We show, in this paper,  that when one has Hamiltonians representing interactions
of multimode radiation fields, i.e., three or more modes, then the dynamical symmetry algebra
of the Hamiltonian becomes a {\it polynomially deformed}  algebra. The deformation is
quadratic for the three mode case and cubic for the four mode case.
The quadratic algebra was discovered by Sklyanin \cite{skl1,skl}, in the context
of statistical physics and field theory \cite{drin}. It was shown  to be the symmetry algebra of a two-dimensional anisotropic harmonic oscillator and the isotropic harmonic oscillator in curved space \cite{le,higgs,tj}. The well-known Higgs algebra, a
cubic algebra, occurs in the study of the dynamical symmetries of
the Coulomb problem in a space of constant curvature\cite{higgs,ze1}.
These algebras have now found a place in quantum optics with the observation that quantum optical Hamiltonians  
describing multiphoton processes have  symmetries which can be described by polynomially deformed
SU(1,1) and SU(2) algebras \cite{kara1}.

A polynomial deformation of a Lie algebra is defined in the following fashion in the
Cartan-Weyl basis,
\be
[H\, , \,E_{\pm}]  = \pm E_{\pm}\,\,\,\,\,, \,\,\,\,
[E_+\, , \,E_-]  =  f(H)\,\,\,\,,
\ee
where $f(H)$ is a {\it polynomial} function of $H$.
The corresponding Casimir can be written in the form \cite{rocek}, 
\bea
C&=&E_- E_+ \,+\,g(H)\,\,\,, \n\\
 &=& E_+E_-\,+\,g(H-1)\,\,\,,
\eea
where,
\be
 f(H)=g(H)-g(H-1).
\ee
The form of $ g(H)$ can be determined up to the addition of
a constant. The eigenstates are characterized by the values of the Casimir
operator and the Cartan
subalgebra $H$.

In particular, a polynomial deformation of $SL(2,R)$ is of the form
$N_0=J_0 $, $ N_+=F(J_0,J)J_+ $, $ N_-=F(J_0,J)J_- $,  where the $J_i$'s
are the ordinary $SL(2,R)$ generators. The commutation relations are
$[N_0,N_{\pm}]=\pm N_{\pm}$ and $[N_+,N_-]=F(N_0)$ \cite{ rocek,curt}.
 When $F(N_0)$ is quadratic in $N_0$ the algebra is called a quadratic algebra
and if it is cubic in $N_0$  the "Higgs" algebra results.

As an example of occurrence of non-linear algebras in quantum optics,
consider  the triboson Hamiltonian
\be
H= \omega _{a} a^{\dag}a + \omega _{b} b^{\dag}b + \omega _{c} c^{\dag}c
+ \kappa a b^{\dag}c^{\dag} + \kappa^{*} a^{\dag} bc .
\ee
Physically, for this Hamlitonian, $a$,$b$ and $c$ represent the pump, signal and idler
modes for parametric amplification and  the idler, pump and signal modes
in frequency conversion.
 Raman and Brillouin scattering can be described by $H$, if $a$, $b$
and $c$ represent input, vibrational and Stokes modes for a Stokes process and
anti-Stokes, input and vibrational modes for an anti-Stokes process.
It also represents the interaction of N identical two-level atoms with a
single mode radiation field.
This has been considered by many authors {\cite {mehta, russ, kara,brif}
 by ordinary linear Lie
algebraic methods leading to  approximate results for specific cases.
Infinite dimensional Lie
algebraic techinques have also been attempted and the physics has been extracted by a truncation of these algebras,
hence the results obtained have again been approximate, with a number of assumptions \cite{brif}.

In this paper, we show that  this Hamiltonian and its generalizations
have a non-linear algebra as its dynamical symmetry algebra
and the construction of the coherent states is straightforward using the representation
theory of these algebras.
Furthermore, all three types of coherent states can be constructed on the basis
of our method.
We shall show that this Hamiltonian is formed by operators which
obey a finite quadratic polynomial deformation of $SL(2,R)$ and the
construction of CS for this Hamiltonian is a fairly straightforward
process based on the undeformed algebra.

For the Hamiltonian given in equation (4), let $a$ represent a pump system and $b$ and $c$ represent
the signal and idler variables. The interaction Hamiltonian between
the pump and signal-idler subsystem is given by
\be
H_{int} = \kappa a b^{\dag}c^{\dag} + \kappa^{*} a^{\dag} bc .
\ee
Energy conservation requires that
$\omega_{a} = \omega_{b} + \omega_{c}$.
If the signal and idler frequencies are equal then
$\omega_{b}=\fr{\omega _{a}}{2}$ and $\omega _{c} =\fr{\omega _{b}}{2}$ with, 
 $H_0^{free} = \omega _{a}(a^{\dag}a  + \fr{b^{\dag}b + c^{\dag}c}{2})$.

The generators of the polynomial quadratic algebra are defined by the operators
\be
J_0 
    =\fr{1}{2} (a^{\dag}a- K_{0})
\ee
\be
J_{-}= a\, b^{\dag}c^{\dag}            
     = a\, K_{+}
\ee
\be
J_{+}= a^{\dag}bc                       
     = a^{\dag} K_{-}
\ee
where $K_{0}$, $ K_{-}$ and $K_{+}$ form  SU(1,1) generators.
The algebra closes only  if we define an additional conserved quantity  $H_0$ given by :
\be
H_0
  = \fr{a^{\dag}a + K_{0}}{2}.
\ee
Since $H_0$ is related to $H_0^{free}$ through $H_0^{free}=2 H_{0}-\frac{1}{2}$, we see that physically this condition is satisfied.
In fact, $H_{0}$ can also be related to the Manley-Rowe invariants of the system.
The algebra is given by:
\be
[J_{+}, J_{-}] = -3J^{2}_{0} +(2 H_{0} -1) J_{0}  -  C_{bc}(K_{0})
+H_{0}(H_{0} + 1)
\ee
Where $C_{bc}=\frac{1}{4}-\frac{(b^{\dag}b-c^{\dag}c)^2}{4}=\frac{1-Q^2}{4}$  is the Casimir operator for the idler-signal system , for which Q is a conserved quantity.
For the case special  $b^{\dag}b-c^{\dag}c=0$ , $C_{bc}=\frac{1}{4}.$

The two commuting generators are then $H_{0}$ and $J_{0}$ and
a general eigenstate of the system is labelled by the quantum numbers corresponding to their eigenvalues and is given by $|h_0,j_0>$.
Similarly,  the symmetry algebra  for four photon processes is a Higgs algebra.

For general  multiphoton Hamiltonians:
\begin{equation}
H=H_{0} + \kappa(a_0)^m(a_1^{\dagger})^n+c.c
\end{equation}
we can define $N_0,N_-,N_+$ in an analogous way\\
\bea
N_+&=&a_0^m(a_1^{\dagger})^n \nonumber\\
N_-&=&a_1^n(a_0^{\dagger})^m \nonumber\\
N_0&=&\frac{1}{m+n}(a_1^{\dagger}a_1-a_0^{\dagger}a_0)
\eea
and show that we get n-dimensional polynomial algebras as the symmetry algebra if
 $H_{0} = \frac{1}{m+n}(a^{\dagger}_0a_0 + a^{\dagger}_1 a_1)$
 is conserved.

Similarly  n-photon Dicke Models with Hamiltonians of the form:
\begin{equation}
H=H_{0} +\kappa\sum_i\sigma_-(i)(a_1^{\dagger})^n +
\kappa^*\sum_i\sigma_+(i)(a_1)^n 
\end{equation}
with
\begin{eqnarray}
N_0&=&\sum_i^{n}\sigma_0(i) - a^{\dagger}_1a_1  \nonumber \\
N_-&=&\sum_i^{n}\sigma_-(i)(a_1^{\dagger})^n  \nonumber \\
N_+&=&\sum_i\sigma_+(i)(a_1)^n 
\end{eqnarray}
 satisfying  a polynomial Lie Algebra of order n if
$H_{0} =\epsilon \sum_{i}^n\sigma_0(i) + w_1a^{\dagger}_1a_1$ is conserved.

 We present a unified approach for finding the
coherent states (CS) of these  algebras . 
Apart from its application to quantum optics, the method of construction
presented here is quite general and will greatly facilitate the physical applications of
these algebras to many quantum mechanical problems.
This method is a generalisation to non-linear algebras  of the procedure 
for constructing multiphoton states outlined in reference \cite{shanta}.
For ordinary Lie algebras, the construction of the CS for the non-compact cases, was shown to be
 a two
step procedure . First,  the canonical conjugate of the  lowering operator were found
and the CS of these  algebras were obtained by
the action of the exponential of the respective conjugate operators on the
vacuum  \cite{shanta,sc,pani}. This method was in complete parallel to the one used for
constructing the coherent states for harmonic oscillator algebras.
 Another CS, dual to the first one, 
 naturally follows from the above construction.  Here,  we generalise the above construction to non-linear algebras and
provide a mapping between the deformed algebras  and their undeformed
counterparts. This mapping is utilized to find the CS in the
Perelomov sense\cite{perv}. Apart from obtaining the known CS of the
$SU(1,1)$ algebra, we construct the CS for the quadratic and cubic 
 polynomial algebras.
Other coherent states in the literature which are essentially special
cases of this construction are the `f-oscillator
states' \cite{sudarshan}
and the non-linear states  $ f(N_0)a|\lambda>=\lambda|\lambda>$,\cite{vogel} 
 which have been shown to be useful for the trapped ion problem.
While these are non-linear {\it harmonic oscillator} coherent states, the CS that we construct
may be called {\it non linear} SU(1,1) (or SU(2)) coherent states.
These states would give a multi-mode generalization of the type  $f(n_a,n_b)a^{n}b^{m}|\lambda>=
\lambda|\lambda>$, as {\it one} of the possible coherent states. Thus our construction encompasses existing non-linear states and allows  for the construction of new physical states.
One such state, for example, is the case n=1 and m=1, which is a two-mode realization of the non-linear coherent states.
 Our method is quite general and encompasses q-deformations \cite{sun}
of linear Lie algebras. In this work, we concentrate on {\it finite, polynomial} SU(2)
and SU(1,1) algebras, in view of their importance in quantum optics.

\section{Construction of Coherent States of Non-Linear Algebras:Formalism.}
 Having seen that polynomially deformed  algebras occur in a large class of systems,
  we now give the formalism  
for the 
 construction of coherent states of these algebras.
 For the purpose of clarity, we start with Lie Algebras and then 
 extend the method to the deformed algebras in a straightforward way. 
In the next section, we shall show how this formalism can be used to explicitly construct the coherent states
for application to mutiphoton processes.

We introduce our method by first considering SU(1,1) for which the generators satisfy the commutation relations
\be
   [K_+\, , \,K_-]  =  -2K_0  
\,\,,\,\,[K_{0} \,,\,K_{\pm}] =  \pm K_{\pm}\,\,\,\,.
\ee
Thus for this case
one finds, $f(K_0)= -2K_0$ and $g(K_0)=-K_0 (K_0 + 1)$. The quadratic
Casimir operator is given by $C=K_- K_+ + g(K_0) =K_-K_+ - K_0(K_0+1)$.
$\tilde{K_{+}}$, the canonical conjugate of $K_- $, satisfying
\be
[K_- \,,\,\tilde{K_+}]=1\,\,\,\,,
\ee
can be written in the form,
\be
\tilde{K_+}=K_+ F(C,K_0)\,\,\,\,.
\ee
Eq.(17) then yields,
\be
F(C,K_0)K_-K_+ - F(C,K_0-1)K_+K_- = 1\,\,\,\,;
\ee
making use of the Casimir operator relation given earlier, one can solve
for $F(C,K_0)$ in the form,
\be
F(C,K_0)=\frac{K_0 + \alpha}{C+K_0(K_0 +1)}\,\,\,\,.
\ee
The constant, arbitrary, parameter $\alpha$ in F can be determined by
demanding that Eq.$(17)$ be valid in the entire Hilbert space.

For the purpose of illustration, we demonstrate our method, 
by using the one oscillator
realization of the $SU(1,1)$ generators $K_-=a^2$,$K_+=a^{\dag^2}$, $K_0=\frac{1}{4}(a a^{\dag} + a^{\dag}a)$. Since the
coherent states of this realization have been studied extensively in the literature \cite{trif,gs,cs}, this provides a
good testing ground for our method.
The ground states defined by $K_- \mid 0>=\frac{1}{2} a^2 \mid 0>=0$, are, 
  $\mid 0>$ and $\mid 1>=a^{\dag}\mid 0>$, in terms of the oscillator
Fock space. 
\be
K_0 \mid 0> =\frac{1}{4}(2a^{\dagger}a + 1) \mid 0> =\frac{1}{4} \mid 0>\,\,\,\,,
\ee
and
\be
C \mid 0>=\frac{3}{16}\mid 0>\,\,\,\,.
\ee
Thus $[K_-\,,\,\tilde{K_+}]\mid 0>=K_-\tilde{K_+}\mid 0>$ yields $\alpha =\frac{3}{4}$.

Similarly, for the other ground state $|1>$,
\be
[K_-\,,\,\tilde{K_+}] \mid 1>=\mid 1>\,\,\,\,,
\ee
leads to $\alpha = \frac{1}{4}$.\\
Hence, there are two disjoint sectors characterized by the $\alpha$ values
$\frac{3}{4}$ and $\frac{1}{4}$, respectively. These results match
identically with the earlier known ones \cite{shanta}, once we rewrite
$F$ as,
\be
F(C,K_0)= \frac{K_0 + \alpha}{C + K_0(K_0 + 1)}\,\,\,\,,
\ee
\be
 = \frac{K_0 + \alpha}{K_- K_+}\,\,\,\,.
\ee
The unnormalized coherent state $\mid\beta>$, which is the annihilation
operator eigenstate, i.e, $K_-\mid \beta >=\beta \mid \beta >$, is given
in the vacuum sector $|0>$ by
\be
\mid \beta> = e^{\beta \tilde{K^+}} \mid 0>\,\,\,\,.
\ee
For the vacuum sector $|1>$, where $\alpha=\frac{1}{4}$, a similar construction holds.
These states, which provide a realization of the Cat states\cite{hill},  play a
prominent role in  quantum measurement theory.
The canonical conjugate  $\tilde{K_{+}}$ such that:
\be
[\tilde{K_{+}^{\dagger}}\,,\,K_+]=1\,\,\,\,.
\ee
can be constructed, as in ref. \cite{shanta} 
from this, one can find the eigenstate of $\tilde{K^{\dagger}_{+}}$
operator, in the form,
\be
\mid \gamma>=e^{\gamma K_+}\mid 0>\,\,\,\,.
\ee
This CS, after proper normalization, is the well-known Yuen (squeezed) state: $e^{\mu a^{\dag^{2}}-\mu^{*}a^2}$, with $\gamma=\frac{\mu}{|\mu|}tanh(|\mu|)$ \cite{perv,yuen}.
Our construction can be easily generalized to various other realizations
of the $SU(1,1)$ algebra, such as the two mode realization, where the corresponding states are the Pair coherent and Perelomov(Cave-Schumaker) states. \\

We now extend the above procedure to the quadratic algebra, which is the relevant algebra in considering the
coherent states of trilinear boson Hamiltonians \cite{junk}.
The algebra is given by:
\be
[J_0\,,\,J_{\pm}]=\pm J_{\pm}\,\,\,,\,\,\,[J_+ \,,\,J_-]=
+(2 H_{0} -1)J_0 -3 J_{0}^{2} - \frac{1-q^2}{4}+H_{0}(H_{0} + 1) \,\,\,\,.
\ee
where the  positive or negative sign of ($2 H_{0} - 1$) determines whether the algebra is a quadratic {\it deformation} of $SU(2)$ or $SU(1,1)$ respectively.
  
In this case,
\bea
f_1(J_0)&=&(2 H_{0} - 1)J_0 - 3 J_{0}^{2} - \frac{1-Q^2}{4}+ H_{0}(H_{0} + 1) \nonumber \\
&=&g_1(J_0)-g_1(J_0 -1),
\eea
where
\be
g_1(J_0)=  J_0 [H_{0}(H_{0} +1)- \fr{1-Q^2}{4} +(H_{0} -\fr{1}{2})(J_{0}+1)]
- J_0(J_0+1)(J_0 + \frac{1}{2})\,\,\,\,.
\ee
In this case we have three different vacua,
$\md h_{0},h_{0} + \fr{q}{2} >$, $ \md h_{0},h_{0} - \fr{q}{2} >$ and
$ \md h_{0},-h_{0} >$ where $h_{0}$ is the eigenvalue of the operator $H_{0}$
and q is the eigenvalue of $ Q= b^{\dag}b - c^{\dag}c $.

This is a special case of the general quadratic algebra:
\be
[N_0\,,\,N_{\pm}]=\pm N_{\pm}\,\,\,,\,\,\,[N_+ \,,\,N_-]=\pm 2bN_0 + a N_{0}^{2}+c\,\,\,\,.
\ee

In this case, $f_1(N_0)=\pm 2bN_0 + a N_{0}^2+c=g_1(N_0)-g_1(N_0 -1)$.
with $ g_{1}(N_{0})=\fr{a}{3} N_{0}(N_{0} +1)(N_{0}+ \fr{1}{2}) + N_{0}(c \pm b(N_{0}+1))$.

The representation theory of the quadratic algebra has been studied in the
literature\cite{rocek},\cite{vin}. It shows a rich structure depending on the values
of `a'. In the non-compact case, i.e, for polynomial deformations of $SU(1,1)$,
the unitary irreducible representations (UIREP) are either bounded below or
above, we can construct the canonical conjugate $\tilde{N_{+}}$ of $N_-$
such that $[N_-\,,\,\tilde{N_+}]=1$. It is given
by $\tilde{N_+}=N_+F_{1}(C,N_0)$, with
\be
F_{1}(C,N_0)=\frac{N_0 + \delta}{C(N_{0})- \fr{a}{3} N_{0}(N_{0} -1)
(N_{0} + \fr{1}{2} ) - N_{0}(c \pm b(N_{0} + 1))} \,\,\,\,.
\ee
As can be seen easily, in the case of the finite dimensional UIREP,  
 $\tilde{N_+}$ is not well defined since $F_1(C,N_0)$ diverges on the
highest state. The values of $\delta$ can be fixed
by demanding that the relation, $[N_-\,,\,\tilde{N_+}]=1$, holds in the
vacuum sector $\mid v_{i} >$,  where , $\mid v_{i} >$'s are annihilated by $N_-$.
This gives $N_-\tilde{N_+}\mid v_{i}>=\mid v_{i}>$, which leads to
$(N_0+\delta)\mid v_{i}>=\mid v_{i}>$. The value of the Casimir operator,
$C=N_-N_+ + g_1(N_0)$, can then be calculated.
Hence, the unnormalized coherent state $\md \alpha >$, such that  $N_-\md \alpha>=
 \alpha \md \alpha >$ is given by $e^{\alpha \tilde{N_+}}\md v_{i}>$.
 We can define the canonical conjugate of $N_{+}$ by $[\tilde{N_{+}^{\dagger}}\,,\,N_+]=1$.
The other coherent state
is $\md \gamma >=e^{\gamma N_+}\mid \tilde{v}_i>$ , where
  $\tilde{N_{+}^{\dagger}}|\tilde{v}_i>=0$.
 Depending on whether the
UIREP is infinite or finite dimensional, this quadratic algebra can
also be mapped onto  the $SU(1,1)$ and $SU(2)$ algebras, respectively. Leaving
aside the commutators not affected by this mapping, one gets,
\be
[N_+\,,\,\bar{N_-}] =- 2bN_0\,\,\,\,;
\ee
where $b=1$ corresponds to the $SU(1,1)$ and $b=-1$ gives the $SU(2)$ algebra.
Explicitly,
\be
\bar{N_-} = N_- G_1(C,N_0)\,\,\,\,,
\ee
and
\be
G_1(C,N_0) = \frac{(N_{0}^2 - N_0)b+\epsilon}{C-g_1(N_0 - 1)}\,\,\,\,,
\ee
$\epsilon$ being an arbitrary constant. One can immediately construct CS in the Perelomov sense (see page 73-74 in ref\cite{perv}) as
$|\xi>= U\mid v_{i}>$, where $U=e^{\eta N_+ -\eta^{\ast} \bar{N_-}}$, with $\xi=\frac{\eta}{|\eta|}tanh(|\eta|)$. 
For the compact case, the CS are analogous to the spin and atomic coherent
states\cite{atcs,spin}.

The cubic algebra, which is also popularly known as the Higgs algebra in the
literature, appears in the study of the Coulomb problem
 in a curved space\cite{higgs} and in quantum optics for quadrilinear boson Hamiltonians.
The generators satisfy,
\be
[M_{0}\,,\,M_{\pm}] = \pm M_{\pm}\,\,\,,\,\,[M_{+}\,,\,M_{-}] = 2cM_0\,+\,4hM_{0}^3\,\,\,\,,
\ee
where, $f_2(M_0) = 2cM_0\,+\,4hM_{0}^3 = g_2(M_0)-g_2(M_0 - 1)$, and
\be
g_2(M_0) = cM_0(M_0 + 1)\,+\,hM_{0}^2(M_0 + 1)^2\,\,\,\,.
\ee
Analysis of its representation theory yields a variety of UIREP's, both finite
and infinite dimensional, depending on the values of the parameters
$c$ and $h$ \cite{zed}. In the non-compact case the canonical conjugate is given by,
\be
\tilde{M_+} = M_+ F_{2}(C,M_0)\,\,\,\,,
\ee
where,
\be
F_{2}(C,M_0) = \frac{M_0 + \zeta}{C-cM_0(M_0+1)-hM_{0}^2(M_0 + 1)^2}\,\,\,\,.
\ee
As before, the annihilation operator eigenstate is given by
\be
\mid \rho>_i= e^{\rho\tilde{M_+}}\mid p_{i}>\,\,\,\,,
\ee
where, $\mid p_{i}>$ are the states annihilated by $M_-$. Like the previous
cases, the dual algebra yields another coherent state.  
This algebra can also be mapped in to $SU(1,1)$ and $SU(2)$ algebras, as
has been done for the quadratic case:
\be
[M_+\,,\,\bar{M_-}]=-2dM_0\,\,\,\,,
\ee
where, $d=1$ and $d=-1$ correspond to the $SU(1,1)$ and $SU(2)$ algebras respectively.
Here,
\be
\bar{M_-}=M_-G_2(C,M_0)\,\,\,\,,
\ee
where,
\be
G_2(C,M_0)=\frac{(M_{0}^2-M_0)d+\sigma}{C-g_2(M_0-1)}\,\,\,\,,
\ee
$\sigma$ being a constant.
The coherent state in the Perelomov sense is then  $ \md \zeta >=U\mid p_{i}>$, where,
$U=e^{\zeta M_+ -\zeta^{\ast} \bar{M_-}}$.
In earlier works on non-linear algebras, the generators of the deformed
algebra have been written in terms of the undeformed ones\cite{rocek,curt}.
However, in our approach the undeformed $SU(1,1)$ and $SU(2)$ generators
are constructed from the deformed generators.


\section{Explicit Construction of the Coherent States for Physical Application.}

We now give an outline of the method of explicit construction of coherent states
for general multiphoton processes for which the generators satisfy the algebra \cite{bon}:
$[N_0\,,N_{\pm}]=\pm N_{\pm}$ and $[N_+\,N_-]=g(N_0) -g(N_0 -1) $

The action on eigenstates of $N_0$ is given by
\be
N_0 \md j,m>=(j+m) \md j,m>
\ee
\be
N_+ \md j,m>=\sq{C(j)-g(j+m)} \md j,m+1>
\ee
\be
N_- \md j,m>=\sq{C(j)-g(j+m-1)} \md j,m-1>
\ee
where $ C(j)=g(j-1) $.\\
Depending on the order of the polynomial algebra n, there will be n+1 degenerate states
annihilated by $N_-$. We denote these as $|j,0>_i$. For each, the vaule of $\delta=\delta_i$ is appropriately chosen as shown earlier.

The coherent state is given by
\bea
\md \al >&=& A e^{\al \tl{N_+}} \md j,0>_i \nonumber \\
&=&A\sm \fr{\al ^n }{n!} (N_+)^n \fr{N_0 +\delta_{i} }{g(j-1)-g(N_0)}
\ld \fr{N_0+n-1+\delta_{i} }{g(j-1)-g(N_0 +n-1)} \md j,0> \nonumber \\
&=&A\sm \al ^n \fr{1}{\sq{(g(j-1)-g(j)) \ld (g(j-1)-g(j+n-1))}} \md j,n>
\eea
`A' being the normalization constant.

A discussion of coherent states is incomplete without showing that these states do give a resolution
of the identity and that they are overcomplete. From the resolution of the identity we have:
\be
\int d\sigma(\alpha^{*},\alpha)\, |\alpha\rangle\langle\alpha| = {\bf 1} 
\ee
Within the polar decomposition
ansatz 
\be
d\sigma(\alpha^{*},\alpha) = \sigma(r) d\theta r dr 
\ee
with $r=|\alpha|$ and an yet unknown positive density $\sigma$  which provides the measure.
For the general case we have:
\be
2\pi\int_{0}^{\infty} dr\, \sigma(r) \, r^{2n+1} = A (g(j-1)-g(j)).....(g(j-1)-g(j+n-1))
\ee
For the various cases the substitution of the explicit value of g(j) then reduces the expression on the R.H.S
to a rational function of  Gamma Functions and the measure $\sigma$ can be found by an inverse Mellin transform.
For the general case the measure is a Meijer's G-function.
The fact that these states are overcomplete (i.e; $<\beta|\alpha>\ne 0$) can be shown for explicit examples.
This we shall later show in the case of a quadratic algebra.

We now construct the state explicitly for purposes of application.
First we  show that, this method  indeed, gives us the well known SU(1,1) Barut-Girardello (pair coherent) states for SU(1,1) in the familair form \cite{Barut}.
The action on Hilbert Space of the generators is given in the original BG representation by:
\be
K_0 \md \ph >=(-\ph + m)\, \md \ph,m >, \\
\ee
\be
K_+ \md \ph,m >= \fr{1}{\sq{2}} \sq{(m+1)(-2\ph + m)}\, \md \ph,m+1>, \\
\ee
\be
K_-  \md \ph,m>= \fr{1}{\sq{2}} \sq{m( -2 \ph + m-1)}\, \md \ph , m-1>. \\
\ee
There are two vacuua annihilated by $K_-$, they correspond to $\md \ph, 0>$ and $\md \ph, 2\ph+1>$.
The  coherent state $ \md \al > $ constructed on the vacuum $\md \ph, 0>$ gives us, $\delta=\ph +1$, so that
$(N_0+\delta )\md \ph,0 > = \md \ph,0>$ and the resulting coherent state is:
 
\be
\md \al >=e^{\al \tilde{K_+}} \md \ph,0> \\
\ee
where, $ [K_- \,,\,\tilde{K_+}]=1 $ and $ \tilde{K_+} =K_+ F(C,K_0)$ with
\be
F(C,K_0)=\fr{K_0 +\delta}{C-g(K_0)} =\fr{K_0+\delta}{C+ \fr{1}{2} K_0 (K_0 + 1)}. \\
\ee
Hence
\be
\md \al >=\sm \fr{\al^n }{n!} (K_+ F(C,K_0))^n \md \ph , 0> =\sm \fr{\al^n}{n!} (K_+)^n F(C,K_0) \ld F(C,K_0 +n-1) \md \ph ,0> \\
\ee
substituting the values of $F$ we get:
\bea
|\alpha>&=&A\sm \fr{\al ^n}{n!} \fr{ (K_0+\delta)(K_0+\delta +1)....(K_0+\delta+n-1)}{(-\ph + \fr{1}{2})( -\ph+\frac{3}{2}).......(-\ph +
\fr{n-1}{2})} \,\, (K_+)^n \, \md \ph ,0> \nonumber \\
&=&A\sm (2\al)^n\fr{\gm (-2\ph )}{\gm(n+1) \gm (-2\ph + n)} \fr{\sq{n! (-2\ph
+ n-1)! }}{(\sq{2})^n \sq{\gm (-2\ph )}} \md \ph ,n> \nonumber \\
&=&A \sq{\gm (-2\ph )} \sm (\sq{2} \al )^n \fr{1}{(\gm (n+1)\, \gm (-2\ph +n))^{\frac{1}{2}}} \md \ph ,n>, 
\eea
which is precisely the well-known state of Barut and Girardello upto the normalization coefficient A.
For example for the SU(1,1), $g(j)=-\frac{1}{2}(j)(j+1)$ then the right hand side becomes in the BG representation
\be
2\pi\int_{0}^{\infty} dr\, \sigma(r) \, r^{2n+1} =A \frac{\Gamma(n+1) \Gamma(-2\phi+n)}{\Gamma(-2\phi)},
\ee
where A is a numerical constant 
and from the inverse Mellin transform, we get $\sigma(r)= A r^{-2\phi +1}K_{\frac{1}{2}+\phi}(2r)$

The second state annihilated by $K_-$ is the state $\md \phi,2\phi +1>$ and this corresponds to $\delta=-\phi$.
 The coherent state is:

\be
|\alpha>=A' \sq{\gm (2\ph )} \sm (\sq{2} \al )^n \fr{1}{\sq{\gm (n+1) \gm (2\ph +n)}} \md \ph ,2\phi +1+n> 
\ee
The third state given by Eq. (29) is :

\be
|\gamma>=B(\gamma)  \sm (\frac{ \gamma}{\sq{2}} )^n \sq{\fr{\gm(n-2\phi)}{\gm (n+1)\,\gm(-2\phi)}} \md \ph ,n> .
\ee
This is the state constructed by Perelomov \cite{perv}, upto a normalisation constant $ B(\gamma)$.

For the quadratic case, we take an illustrative algebra  relevant to the trilinear boson cases described in the introduction.
For convenience we rewrite the three boson algebra as:\\
 $[N_0\,,N_{\pm}]=N_{\pm}$ and $ [N_+\,N_-]=-3N_{0}^{2} +4\epsilon N_0-\epsilon^2 $
with $\epsilon=2 H_0 -1 $. We define
$n=(h_0 + j_0)$, where $h_0$ and $j_0$ are the quantum numbers associated with $H_0$ and $N_0$ respectively. The  state $|n>$ corresponds to the state $|h_0,h_0 + j_0>=|\epsilon,n>$
and the three states  annihilated by $N_-$  are given by $|\epsilon,0 >,|\epsilon, \epsilon - \frac{1}{2}>, |\epsilon,\epsilon+\frac{1}{2}>$.

The action of the operators on eigenfunctions of $N_0$ is given by:
\be
N_0 \md \epsilon, n>=(n ) \md \epsilon, n>,
\ee
\be
N_+ \md \epsilon, n>=\sq{(n+\fr{3}{2} -\epsilon)(n+1)(n+ \fr{1}{2} -\epsilon)} \md \epsilon, n+1>,
\ee
\be
N_- \md \epsilon ,n>=\sq{(n-\fr{1}{2} -\epsilon)n(n+\fr{1}{2} -\epsilon)} \md \epsilon, n-1 >.
\ee
We give the explicit construction  of the coherent state for the case $|v_i>=|\epsilon,0>$ , for which $\delta=1$.
 Suitable choices of  $\delta$  will give the other two coherent states.
Here $ g(N_0)=-(N_0+\frac{3}{2} -\epsilon)(N_0 )(N_0 +\frac{1}{2}-\epsilon) $ and
$g(N_0)-g(N_0-1)=3N_{0}^{2} -4\epsilon N_0+\epsilon^2 $.

From our construction the CS is:
\be
\md \al >=e^{\al \tl{N_+}} \md \epsilon, 0>=\sm \fr{\al^n }{n!} (\tl{N_+})^n \md \epsilon, 0>
\ee
Thus:
\be
|\alpha>=A\sm \fr{\al^n }{n!} (N_+ F(N_0,C))^n \md 0>
\ee
Constructing the $F's$ from  $g(N_0)$ we get:
\bea
|\alpha>&=&A\sm \fr{\al^n }{n!} (N_+)^n F(N_0)F(N_0 +1) \ld F(N_0 +n-1) \md 0> \nonumber \\
&=&A\sm \fr{\al^n }{n!} (N_+)^n \fr{N_0 +\delta}{(N_0 -\epsilon)(N_0 )
(N_0 +1-\epsilon)} \ld \fr{N_0 +n-1+\delta }{(N_0 +n-1-\epsilon)(N_0 +n)
(N_0 +n-\epsilon)} \md \epsilon, 0> \nonumber \\
&=&A\sm \al ^n  \fr{(-\fr{1}{2} -\epsilon)!(\fr{1}{2} -\epsilon)!}{(n-\fr{1}{2} -\epsilon)!
n!(n+\fr{1}{2} -\epsilon)!} (N_+)^n \md \epsilon, 0> \nonumber \\
&=&A\sq{\gm (\fr{1}{2} -\epsilon) \gm (\fr{3}{2} -\epsilon)} \sm
\fr{\alpha^n}{\sq{\gm (n+\fr{1}{2} -\epsilon) \gm (n+1) \gm (n+\fr{3}{2} -\epsilon)}}
\md \epsilon, n>,
\eea
$A$ is the normalization coefficient, which can be easily determined
to be $\frac{1}{(^0F_2(\frac{1}{2}-\epsilon,\frac{3}{2}-\epsilon,|\alpha|^2))^{\frac{1}{2}}}$.
These set of states  can be shown to be overcomplete:
\be
|<\beta|\alpha>|^2= \frac{^0F_2(\frac{1}{2}-\epsilon,\frac{3}{2}-\epsilon,\alpha\beta^{*})}
{(^0F_2(\frac{1}{2}-\epsilon,\frac{3}{2}-\epsilon,|\alpha|^2))^{\frac{1}{2}})
(^0F_2(\frac{1}{2}-\epsilon,\frac{3}{2}-\epsilon,|\beta|^2))^{\frac{1}{2}}}.
\ee

The completeness relation is given by
\begin{equation}
\int_{0}^{\infty} dr\, \sigma(r) \, r^{2n+1} = \Gamma(n+1)
\, \frac{\Gamma(\frac{1}{2}-\epsilon+n)
\Gamma(\frac{3}{2}-\epsilon+n)}
{\Gamma(\frac{1}{2}-\epsilon)\Gamma(\frac{3}{2}-\epsilon)}
\end{equation}
and $\sigma(r)$ can be determined to be a confluent hypergeometric function from the inverse Mellin transformation formula.
The resolution of the identity can thus be obtained.
\be
\sigma(r)=\fr{1}{\Gamma(\frac{1}{2}-\epsilon)(\Gamma(\frac{3}{2}-\epsilon )}
G^{3\,0}_{0\,3} (r\md^{0}_{0,-\fr{1}{2} - \epsilon ,\fr{1}{2} -\epsilon} ),
\ee
Where $ G^{3\,0}_{0\,3}(x)$ is a Meijer's G-function.

The other two coherent states based on the two vacuua, $|\epsilon, \epsilon+\frac{1}{2}>$ and $|\epsilon, \epsilon-\frac{1}{2}>$
can similarly be constructed by chosing $\delta=\frac{3}{2}-\epsilon$ and $\delta=\frac{1}{2}-\epsilon$.

The state corresponding to the Perelomov state is :
\be
|\gamma>=B'(\gamma)  \sm ( \gamma)^n \sq{\fr{\gm(n+\frac{3}{2}-\epsilon)\gm(n+\frac{1}{2}-\epsilon)}{\gm (n+1)\,\gm(\frac{1}{2}-\epsilon) \gm(\frac{3}{2}-\epsilon)}} \md \ph ,n> .
\ee

The normalisation constant can be calculated easily and using a method similar to the one used for Eq.(68), the overcompleteness of these states and the resolution of the identity can also be easily obtained.


\section{Conclusion}

To conclude, we have found a general method for constructing the coherent
states for various polynomially deformed algebras for quantum optical
systems, whose dynamics are governed by multilinear boson Hamiltonians.
Since our method is algebraic and
relies on the group structure
of well-known algebras, the precise nature of the non-classical behaviour
of these CS can be easily inferred from our construction. It will be of particular
interest to see the time development of the system and
the role of the deformation parameters in the physical system described in
the text. For a system initially in a coherent state,
it is fairly straightforward to calculate the time evolution of the system exactly using the methods of reference \cite{gs}.
Since many of these  algebras are related to
quantum mechanical problems with non-quadratic, non-linear Hamiltonians,
 a detailed study of
the properties of the CS associated
with  non-linear and deformed algebras is of physical relevance \cite{vin,deb}.
This is the subject of our current and future work \cite{sun}.

The authors take the pleasure to thank Prof. S. Chaturvedi and  Prof. C. Mukku for stimulating
conversations. VSK acknowledges useful discussions with Mr. N. Gurappa.


\end{document}